# Engineering direct-indirect band gap transition in wurtzite GaAs nanowires through size and uniaxial strain

Andrew Copple,[1] Nathaniel Ralston,[1] Xihong Peng,[2*]

[1]Department of Physics, Arizona State University, Tempe, Arizona 85287, USA
[2]Department of Applied Sciences and Mathematics, Arizona State University, Mesa, Arizona 85212, USA

**ABSTRACT**

Electronic structures of wurtzite GaAs nanowires in the [0001] direction were studied using first-principles calculations. It was found that the band gap of GaAs nanowires experience a direct-to-indirect transition when the diameter of the nanowires is smaller than ~28 Å. For those thin GaAs nanowires with an indirect band gap, it was found that the gap can be tuned to be direct if a moderate external uniaxial strain is applied. Both tensile and compressive strain can trigger the indirect-to-direct gap transition. The critical strains for the gap-transition are determined by the energy crossover of two states in conduction bands.

Keywords: electronic band structure, direct/indirect band gap transition, wurtzite GaAs nanowire, uniaxial strain, quantum confinement

One dimensional nanostructures, such as nanowires, of group III-V semiconductors have drawn extensive research interests in recent years. They are expected to play important roles as functional components[1] in future nanoscale field effect transistors,[2] high efficiency photo detectors,[3,4] light emitting diodes[5], photovoltaic cells,[6] medicine sensors,[7] etc. In particular, GaAs has been considered as a promising channel material for the high speed NMOS beyond Si based technology. GaAs has two different crystal structures zinc blende (ZB) and wurtzite (WZ) phases. In bulk GaAs, ZB phase is energetically more favorable than WZ. In nanoscale, however, WZ phase was observed more often experimentally. Theoretical work,[8] including *ab-initio* calculations, has shown that at small size WZ structure is energetically more favorable,[8-10] and the interface energy may also facilitate the growth of WZ structure.[11] Recent experiments have shown it is possible to grow GaAs nanowires in ZB,[12] WZ,[13] and mixed crystal phases.[14]

While bulk GaAs (both ZB and WZ) has a direct band gap, GaAs nanowires may demonstrate an indirect gap when the diameter of nanowire is sufficiently small.[15,16] This band gap transition could fundamentally alter the electronic properties of nanowires. In addition to size, strain has become a routine factor to engineer band gaps of semiconductors in the field of microelectronics. Researchers have theoretically demonstrated the modulated band gap by external strains in a variety of systems such as pure Si[17] and Ge[18] and Si/Ge Core-shell nanowires.[19] It would be very interesting to investigate strain effects on the band structure of WZ GaAs nanowires and examine if the direct-indirect band gap transition can be engineered for applications.

The *ab-initio* density functional theory (DFT)[20] calculations were carried out using VASP code[21,22]. The DFT local density approximation and the projector-augmented wave potentials[23,24] were used along with plane wave basis sets. The kinetic energy cutoff for the plane wave basis set was chosen to be 300.0 eV. The energy convergence criteria for electronic and ionic iterations are $10^{-4}$ eV and 0.03 eV/Å, respectively. The GaAs nanowires were generated along the [0001] direction (i.e. *z*-axis) from bulk WZ GaAs with different diameters in the wire cross section (see Figure 1). The dangling bonds on the wire surface are saturated by hydrogen atoms.

The Ga 3d, 4s, 4p, As 4s, 4p and H 1s electrons are treated as valence electrons. The reciprocal space of a nanowire is sampled at 1×1×6 using Monkhorst Pack meshes. In band structure calculations, a total of 21 k-points were included along the K vector direction Γ(0, 0, 0) to X (0, 0, 0.5). The initial axial lattice constant in GaAs nanowires is set to be 6.5083 Å, taken from the relaxed lattice constant $c$ of bulk WZ GaAs. In addition to the axial lattice constant, the lateral length of the simulation cell is chosen so that the distance between the wire and its replica (due to periodic boundary conditions) is more than 15 Å to minimize the interactions between the wire and its replica. The axial lattice constant in a WZ GaAs wire is optimized through the technique of total energy minimization. Once the optimized geometry of a wire is obtained, we applied uniaxial tensile/compressive (i.e. positive/negative) strain to the wire by scaling the axial lattice constant of the wire. For a wire under each strain, the lateral $x$ and $y$ coordinates are further optimized through energy minimization. The band gap of a nanowire is defined by the energy difference between the conduction band minimum (CBM) and the valence band maximum (VBM).

Four different sizes of WZ GaAs nanowires in the [0001] direction with a hexagonal cross section were studied and the corresponding diameters are 6.4 Å, 14.3 Å, 22.2 Å and 30.1 Å, as shown in Figure 1. It was found that the optimized axial lattice constant of the nanowires are the same as its bulk lattice constant ($c$=6.5083 Å), except for the smallest wire. For the smallest wire, the lattice constant is 6.4933 Å, implying a very slight contraction of 0.23% compared to the bulk value.

It is known that bulk WZ GaAs has a direct band gap[8, 25] with both VBM and CBM located at Γ. However, our calculated band structures of thin WZ GaAs nanowires demonstrate an indirect band gap, shown in Figure 2(a)-(c). VBM of the three smaller nanowires are all at Γ and CBM are located at the CB valley along the Γ-X direction. This CB valley could be related to M-L valleys of bulk WZ GaAs.[8] For the larger nanowire with a diameter of 30.1 Å, the band structure demonstrates a direct band gap with both VBM and CBM located at Γ, shown in Figure 2(d). The indirect-direct band gap transition in the WZ GaAs nanowires occurs in the size range 22 ~ 30 Å. From Figure 2, one also can find that the band gap of the nanowires[26] increases with the reduction of the wire diameter, which is mainly due to quantum confinement effects.

To further estimate the size of the nanowire in which the indirect-to-direct-gap transition occurs, we plotted energies of three states, namely VBM, conduction band (CB) at Γ and CB at the valley, as a function of the diameter of the wires, presented in Figure 4(a). It's shown that the energy of VBM ($E_{VBM}$) increases while both energies of CB at Γ ($E_{CB-\Gamma}$) and at the valley ($E_{CB-v}$) reduce with increasing size of the nanowires. For the three smaller wires, the energy of CB at the valley ($E_{CB-v}$) is lower than the energy of CB at Γ ($E_{CB-\Gamma}$), indicating an indirect band gap. However, for the largest wire, $E_{CB-\Gamma}$ becomes lower, therefore having a direct band gap. The indirect-to-direct-gap transition size was estimated to be ~ 28 Å, from the crossover of $E_{CB-\Gamma}$ and $E_{CB-v}$ in Figure 4(a). This critical lateral size to trigger the direct-indirect band gap transition in WZ GaAs nanowires is smaller than that of the ZB nanowires (~40 Å) reported by Persson and Xu, using tight-binding calculations.[15]

Effects of uniaxial strain on the band structures of WZ GaAs nanowires were further studied. It was found that the band gap can be significantly tuned by uniaxial strain. For the thin nanowires with an indirect band gap, a suitable uniaxial strain can tune the gap into a direct band gap. Taking an example of the nanowire with a diameter of 14.3 Å, the effect of uniaxial strain on the band structures were presented in Figure 3. Without strain, the wire has an indirect band gap with CBM located at the valley in Figure 3(c). The indirect/direct band gap nature is determined by the values of $E_{CB-\Gamma}$ and $E_{CB-v}$. With a tensile uniaxial strain, both $E_{CB-\Gamma}$ and $E_{CB-v}$ decrease. However, the downward shift of $E_{CB-\Gamma}$ is larger than that of $E_{CB-v}$, as shown in Figure 3(d)-(f). For example, when a tensile strain of 6% was applied to the wire, $E_{CB-\Gamma}$ is lower than $E_{CB-v}$, suggesting a direct band gap. On the other hand, when a compressive strain was applied, $E_{CB-\Gamma}$ experiences a downward shift while $E_{CB-v}$ increases, as shown in Figure 3(a) and (b). The band gap becomes direct when $E_{CB-\Gamma}$ is lower than $E_{CB-v}$, for example, with a -4% strain.

This indirect-to-direct band gap transition manipulated by strain was also observed for the nanowire with a diameter of 22.2 Å. However, the critical strain to trigger the band gap transition for the 22.2 Å nanowire is smaller than that of the 14.3 Å wire. To illustrate this, three energies $E_{VBM}$, $E_{CB-\Gamma}$ and $E_{CB-v}$ are plotted as a function of strain for these two nanowires and presented in Figure 4(b). Generally for both wires, $E_{CB-v}$ changes linearly with strain, decreasing with tensile strain while increasing with compression. However, $E_{CB-\Gamma}$ has its maximized value at 0% strain and drops with both tensile and compressive strain. The crossover of these two energies $E_{CB-\Gamma}$ and $E_{CB-v}$ is the critical strain for the indirect-to-direct band gap transition. It is clear, from Figure 4(b), the critical strains are roughly -2.2% and +5.2% for the nanowire with a diameter of 14.3 Å. For the larger wire with a diameter of 22.2 Å, the critical strain to trigger the indirect-direct gap transition is approximately -0.8% and +2.8%. For the thinnest nanowire with a diameter of 6.4 Å, no indirect-to-direct band gap transition was found within the uniaxial strain of ±6% considered in this work.

As shown in Figure 4(b), the nearly *linear* shift of $E_{CB-v}$ and $E_{VBM}$ with strain was also observed in other semiconducting nanostructures.[17, 18, 27-30] More interesting in Figure 4(b) is the $E_{CB-\Gamma}$ shift with strain, which is maximized at 0% strain and drops at both tensile and compressive uniaxial strains, demonstrating a *non-linear* behavior. To understand this unique behavior which essentially determines the band gap transition, we further explored the detailed wavefunction and electron density of the state CB at Γ with different values of strain. We found that CB at Γ does not correspond to the same state under different strains. It is a result of a competition between two conduction band states (A and B). To illustrate this, we take the wire with a diameter of 14.3Å as an example (note that the following general conclusions are also valid for the larger wire with a diameter of 22.2 Å). Figure 4(c) displays the electron density contour plots of these two competitive states A and B and their energy shifts with strain. Both energy shifts of states A and B show a *linear* behavior with strain, in which the energy of state A increases with tensile strain while state B demonstrates an opposite trend. It's clear that, without strain and under negative strain, the energy of state A is lower than that of state B, thus state A represents the conduction band at Γ. However, with a tensile strain, state A has a higher energy than B, therefore state B is the conduction band.

To further understand the two distinct *linear* trends of state A and B in Figure 4(c), their wavefunctions were projected onto *s*, *p*, and *d* orbitals and it was found that both states are dominated by the *s*-orbital (>90%). However, the bonds along the *z*-direction show distinct bonding and anti-bonding characteristics for state A and B, respectively (see the electron contour plots). For state A, compressive uniaxial strain along the *z*-direction makes these bonding lengths shorter and makes the electron cloud more effectively shared by nuclei, thus decreasing the electron-nuclei Coulomb potential energy.[29] In contrast, tensile uniaxial strain moves electron cloud farther away from the nuclei and increases the Coulomb potential energy. This explains the linear relation of the energy of state A with strain in Figure 4(c). For state B, the anti-bonding characteristics makes the Coulomb energy between the electron and the nuclei less sensitive to uniaxial strain.[29] However, the anti-bonding characteristics suggests that the nodal surfaces of the positive and negative values of the wavefunction are perpendicular to the *z*-direction and a compressive uniaxial strain reduces the distance between the nodal surfaces, therefore kinetic energy associated with the electron transportation between atoms increases.[30-32] In contrast, a tensile strain increases the nodal surfaces thus reducing the associated kinetic energy.

In summary, thin wurtzite GaAs nanowires along the [0001] direction with a diameter up to 30 Å were studied using DFT calculations. It was found that (1) the band gap of the GaAs nanowire increases when the size of the nanowire decreases due to the quantum confinement; (2) the band gap of the WZ GaAs nanowires experiences a transition from direct to indirect when the diameter of the nanowire is smaller than 28 Å; (3) uniaxial strain can significantly tune the band structure of the nanowires; (4) the indirect band gap of the ultrathin nanowire can be tuned to a direct band gap with an appropriate tensile/compressive uniaxial strain; (5) thinner nanowires require a larger critical strain to trigger the indirect-to-direct band gap transition.

The authors thank Arizona State University Advanced Computing Center for providing computing resources (Saguaro Cluster). Fu Tang is acknowledged for the valuable discussions and the critical review of the manuscript.

* To whom correspondence should be addressed.  E-mail: xihong.peng@asu.edu.

**Figure captions**

**Figure 1. Snapshots of studied wurtzite GaAs nanowries. The diameter and composition of each wire is given on the bottom. Dashed rectangle in (e) indicates a unit cell.**

**Figure 2. Band structures of geometrically optimized wurtzite GaAs nanowires with various diameters. Energies are referenced to vacuum level. There is a conduction band valley away from Γ.**

**Figure 3. Band structure of wurtzite GaAs nanowire with a diameter of 14.3 Å under uniaxial strain. Positive and negative values of strain refer to tensile and compressive strain, respectively. Energies are referenced to vacuum level. Indirect-to-direct band gap transition occurs for expansion beyond 4% and compression larger than 2%. $E_{CB-\Gamma}$ and $E_{CB-v}$ are the energies of the conduction band at Γ and at the valley, respectively.**

**Figure 4. The calculated energies for VBM, conduction band at Γ and at the valley, as a function of (a) the wire diameter and (b) uniaxial strain. In (b), the solid symbols and lines are for the nanowire with a diameter of 14.3 Å, while the hollow symbols and dashed lines are for the wire with a diameter of 22.2 Å. (c) The two competitive states (A and B) representing CB at Γ and their energy trends with strain. The nanowires are along the *z*-direction.**

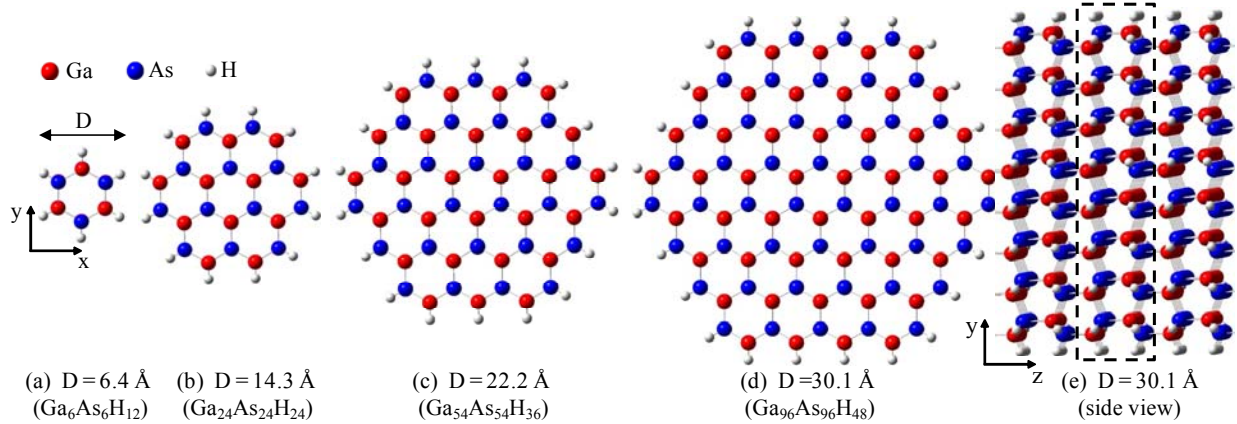

(a) D = 6.4 Å  (b) D = 14.3 Å  (c) D = 22.2 Å  (d) D = 30.1 Å  (e) D = 30.1 Å
($Ga_6As_6H_{12}$)  ($Ga_{24}As_{24}H_{24}$)  ($Ga_{54}As_{54}H_{36}$)  ($Ga_{96}As_{96}H_{48}$)  (side view)

Figure 1. Snapshots of studied wurtzite GaAs nanowries. The diameter and composition of each wire is given on the bottom. Dashed rectangle in (e) indicates a unit cell.

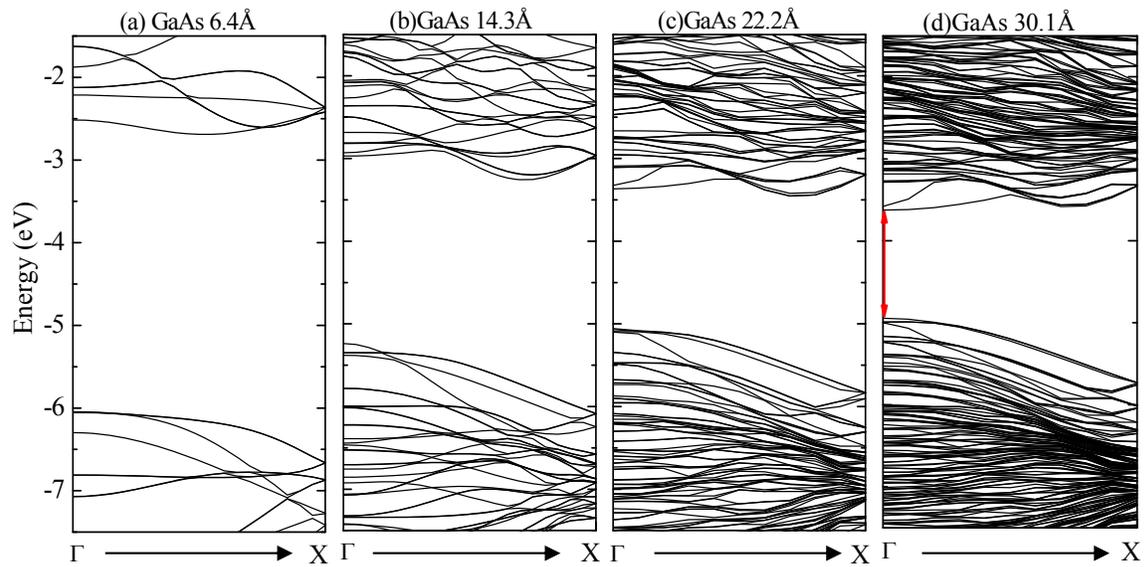

Figure 2. Band structures of geometrically optimized wurtzite GaAs nanowires with various diameters. Energies are referenced to vacuum level. There is a conduction band valley away from Γ.

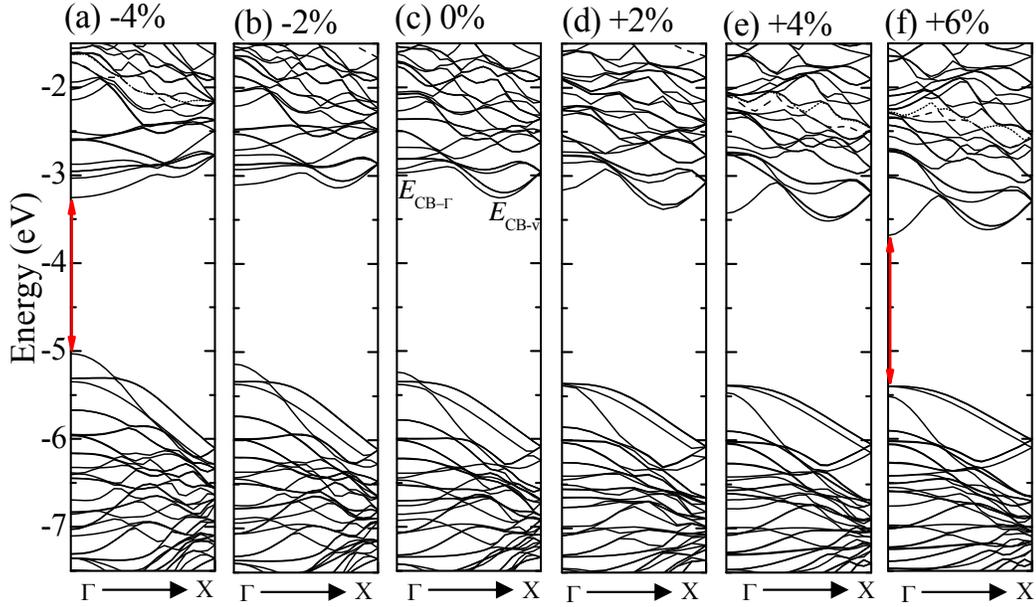

Figure 3. Band structure of wurtzite GaAs nanowire with a diameter of 14.3 Å under uniaxial strain. Positive and negative values of strain refer to tensile and compressive strain, respectively. Energies are referenced to vacuum level. Indirect-to-direct band gap transition occurs for expansion beyond 4% and compression larger than 2%. $E_{CB-\Gamma}$ and $E_{CB-v}$ are the energies of the conduction band at $\Gamma$ and at the valley, respectively.

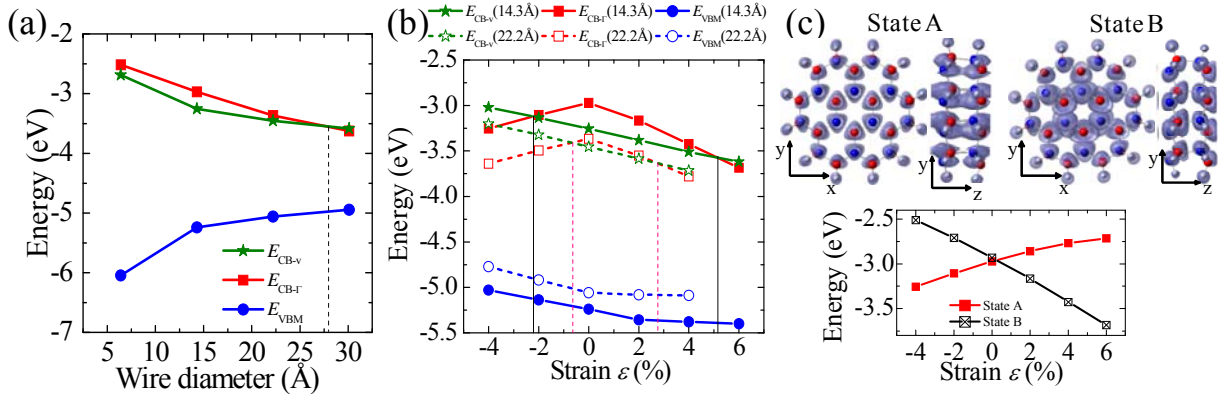

Figure 4. The calculated energies for VBM, conduction band at $\Gamma$ and at the valley, as a function of (a) the wire diameter and (b) uniaxial strain. In (b), the solid symbols and lines are for the nanowire with a diameter of 14.3 Å, while the hollow symbols and dashed lines are for the wire with a diameter of 22.2 Å. (c) The two competitive states (A and B) representing CB at $\Gamma$ and their energy trends with strain. The nanowires are along the $z$-direction.